\documentclass[conference]{IEEEtran}
\IEEEoverridecommandlockouts
\usepackage{cite}
\usepackage{amsmath,amssymb,amsfonts}
\usepackage{algorithmic}
\usepackage{graphicx}
\usepackage{textcomp}
\usepackage{xcolor}
\def\BibTeX{{\rm B\kern-.05em{\sc i\kern-.025em b}\kern-.08em
    T\kern-.1667em\lower.7ex\hbox{E}\kern-.125emX}}

\usepackage{amsmath,amssymb,amsfonts}
\usepackage{algorithmic}
\usepackage{algorithm}
\usepackage{graphicx}
\usepackage{textcomp}
\usepackage{xcolor}
\usepackage{subcaption}
\usepackage{url}
\usepackage{booktabs}
\usepackage[normalem]{ulem}
\usepackage{listings}
\usepackage{xcolor}
\usepackage{tikz}
\usetikzlibrary{decorations.markings}
\usetikzlibrary{shapes.geometric, arrows}
\usetikzlibrary{positioning}
\usetikzlibrary{fit}
\usepackage{pgfplots}
\pgfplotsset{
    scale only axis,
    width=14cm, height=9cm,
    compat=1.18
}
\usetikzlibrary{arrows.meta, positioning, fit, calc, decorations.pathreplacing}

\definecolor{codegreen}{rgb}{0,0.6,0}
\definecolor{codegray}{rgb}{0.5,0.5,0.5}
\definecolor{codepurple}{rgb}{0.58,0,0.82}

\lstdefinestyle{mystyle}{
    commentstyle=\color{codegreen},
    keywordstyle=\color{magenta},
    numberstyle=\tiny\color{codegray},
    stringstyle=\color{codepurple},
    basicstyle=\ttfamily\footnotesize,
    breakatwhitespace=false,
    breaklines=true,
    numbers=left,
    captionpos=b,
    keepspaces=true,
    showtabs=false,
    tabsize=2,
}
\usepackage[mode=buildnew]{standalone}
\usepackage{comment}
\DeclareGraphicsExtensions{.jpg,.pdf}
\begin{document}

\title{A Lock-Free Work-Stealing Algorithm for Bulk Operations\\
\thanks{This project was supported in part by the National Science Foundation grant CMMI-2547883.}
}

\author{\IEEEauthorblockN{Raja Sai Nandhan Yadav Kataru}
 \IEEEauthorblockA{\textit{Department of Computer Science} \\
 \textit{Iowa State University}\\
nandgate@iastate.edu}
\and
\IEEEauthorblockN{Danial Davarnia}
\IEEEauthorblockA{\textit{Edwardson School of Industrial Engineering} \\
\textit{Purdue University}\\
ddavarn@purdue.edu}
\and
\IEEEauthorblockN{Ali Jannesari}
\IEEEauthorblockA{\textit{Department of Computer Science} \\
 \textit{Iowa State University}\\
jannesar@iastate.edu}
}

\maketitle

\begin{abstract}
Work-stealing is a widely used technique for balancing irregular parallel workloads, and most modern runtime systems adopt lock-free work-stealing deques to reduce contention and improve scalability. However, existing algorithms are designed for general-purpose parallel runtimes and often incur overheads that are unnecessary in specialized settings. In this paper, we present a new lock-free work-stealing queue tailored for a master-worker framework used in the parallelization of a mixed-integer programming optimization solver based on decision diagrams. Our design supports native bulk operations, grows without bounds, and assumes at most one owner and one concurrent stealer, thereby eliminating the need for heavy synchronization.

We provide an informal sketch that our queue is linearizable and lock-free under this restricted concurrency model. Benchmarks demonstrate that our implementation achieves constant-latency push performance, remaining stable even as batch size increases, in contrast to existing queues from C++ Taskflow whose latencies grow sharply with batch size. Pop operations perform comparably across all implementations, while our steal operation maintains nearly flat latency across different steal proportions. We also explore an optimized steal variant that reduces latency by up to 3x in practice. Finally, a pseudo workload based on large-graph exploration confirms that all implementations scale linearly. However, we argue that solver workloads with irregular node processing times would further amplify the advantages of our algorithm.  
\end{abstract}


\section{Introduction}

Parallel computing is a computational paradigm in which multiple tasks are executed simultaneously to complete a set of tasks faster than traditional sequential execution. The underlying machine usually contains multiple cores on chip, each running different tasks in parallel. For efficient use of the CPU resources, the processors use different load-balancing strategies to keep the cores busy as evenly as possible. When tasks generate new subtasks during execution, these are often scheduled on the same core that created them, thereby reducing task migration and minimizing the overhead of context switching.

Two types of load-balancing schemes are developed: work-sharing and work-stealing \cite{blumofe}. In work-sharing, cores proactively offload part of their workload whenever new tasks are created. When a core generates new tasks/threads, the scheduler attempts to transfer some of these tasks to other processors, particularly those that appear underutilized. This aims to quickly disseminate work throughout the system, ensuring that idle cores receive new tasks as soon as possible. However, the frequent migration of tasks introduces overhead, since task distribution occurs even when the other cores are already sufficiently occupied. This can lead to higher migration costs, potential contention, and inefficient use of caches.

In work-stealing, by contrast, the responsibility for balancing the workload lies primarily with the idle or underutilized cores. Instead of new tasks being pushed, the idle cores attempt to steal tasks from the queues of busy cores. This reduces the overhead of task migration, since no tasks are moved when all the cores are working. As a result, work-stealing often leads to lower scheduling overhead, and better use of caches. For this reason, many modern parallel programming frameworks (e.g., Cilk \cite{cilk}, Intel TBB \cite{intel2019_tbb}, Java Fork/Join \cite{Java_forkjoin}, and Go's runtime \cite{go_runtime}) adopt work stealing as their primary scheduling strategy. 

\begin{figure}
    \centering
    \includegraphics[width=\columnwidth]{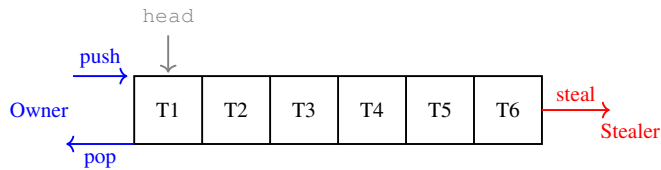}
    \caption[Work-stealing queue]{Classic work-stealing deque model: the owner thread pushes and pops tasks at the top, while idle threads steal tasks from the bottom. A head pointer tracks the top of the deque.}
  \label{fig:classic_deque}
\end{figure}

Work-stealing generally involves maintaining a \textit{deque} (double-ended queue) private to each core. The deque is a data structure that supports popping nodes from both ends. The owner thread pushes and pops nodes from one end, typically at the front, and the other threads may steal from the other end (as shown in Figure \ref{fig:classic_deque}) \cite{blumofe}, \cite{deque_wiki}.  This minimizes the contention between owner and stealer operations while supporting efficient work distribution. However, despite the effectiveness in general purpose parallel runtimes, existing work-stealing algorithms often make assumptions that can introduce unnecessary overhead in specialized settings with different workload characteristics and scheduling policies. 

One specialized setting arises in solving \textit{combinatorial network problems} (CNPs), which are network optimization models that impose combinatorial constraints on certain network elements. CNPs play a central role in applications ranging from transportation and supply chain management to logistics.
A prominent example is the unsplittable flow problem (UFP), a CNP in which the incoming and outgoing arc flows at certain nodes cannot be split or merged \cite{davarnia:ri:ic:ta:2019}. Specifically, each incoming flow at such nodes must be matched with a single outgoing flow. This structure appears in domains such as rail transportation, information infrastructure, and inter-cloud telecommunication design \cite{salemi:da:2023}.

In most applications, the size of the underlying CNP is large, making them extremely difficult to solve due to the complexity of the embedded combinatorial constraints. With the rapid growth of cluster and multicore computing, solving large-scale CNPs increasingly depends on scalable, parallelized solution methods that can efficiently distribute computations across processors. Unfortunately, conventional approaches, such as mixed-integer programming (MIP) and decomposition-based methods, are poorly suited for parallelization. To address this, recent research has introduced techniques based on the concept of \textit{decision diagrams} (DDs), which are compact graphical representations of the optimization problem’s data structure \cite{bergman:ci:va:ho:2016,vanhoeve:2024}. 

\subsection{Decision Diagrams Overview}
Formally, a DD is a directed acyclic graph $D = (V, A, \ell)$ with a node set $V$, arc set $A$, and arc-label function $\ell$. The nodes are partitioned into $n+2$ layers $L_1, \dots, L_{n+2}$, where $n$ is the number of decision variables. The first and last layer contains a single node, the \emph{root} $r \in L_1$ and the \emph{terminal} $t \in L_{n+2}$. The arcs are partitioned into $n$ levels, and each arc $a \in A$ goes from a node in $L_i$ to a node in $L_{i+1}$ with a label $\ell(a) \in D(x_i)$, indicating the assignment to variable $x_i$ at layer $i$. Thus any root-to-terminal path $p=(a^{(1)},\dots,a^{(n)})$ encodes a feasible assignment $x_j=\ell(a^{(j)})$ for $j=1,\dots,n$.

Consider the following \emph{Knapsack} problem as an example.

\begin{equation}
\begin{aligned}
    \max \quad & 8x_1 + 5x_2 + 7x_3 + 6x_4 \\
    \textrm{s.t.} \quad & 3x_1 + 2x_2 + 4x_3 + 6x_4 \leq 7 \\
    & x_i \in \{0,1\}
\end{aligned}
\label{knapsack_eq}
\end{equation}

\begin{figure}
    \centering
    \includegraphics[width=\columnwidth]{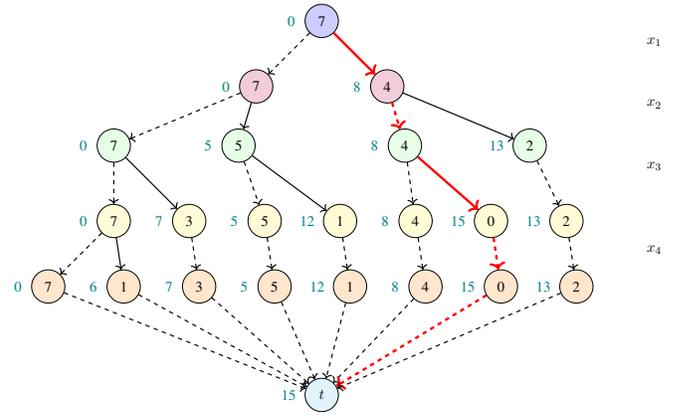}
    \caption{Exact DD for \emph{Knapsack} problem \eqref{knapsack_eq}. The dashed arcs and the solid arcs correspond to the decisions of 0 and 1, respectively. The \emph{longest} path is 15 (highlighted in red), obtained by traversing the path highlighted in red from the root \emph{r} to the terminal \emph{t}.}
    \label{fig:exact_dd}
\end{figure}

Figure \ref{fig:exact_dd} illustrates a DD for \eqref{knapsack_eq}. Each layer corresponds to a variable $x_i$.
The numbers inside the nodes indicate the state (remaining capacity), while the numbers to the left of the nodes denote the cumulative objective value along the path, computed by treating the objective function coefficients as the weights of the corresponding arcs leading to each node. The optimal solution is calculated by the length of the \emph{longest} path from $r$ to $t$.

\begin{figure}
    \centering
    \includegraphics[width=\columnwidth]{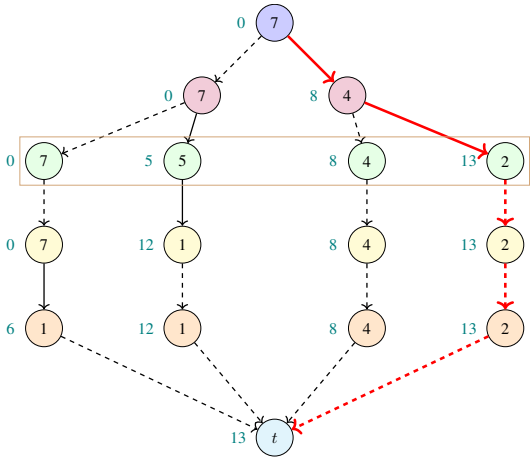}
    \caption{a Restricted DD for \emph{Knapsack} problem \eqref{knapsack_eq}. The dashed arcs and the solid arcs correspond to the decisions of 0 and 1, respectively, for the decision variable. The \emph{longest} path is 13 (highlighted in red), obtained by traversing the path highlighted in red from the root \emph{r} to the terminal \emph{t}.}
    \label{fig:restricted_dd}
\end{figure}

\begin{figure}
    \centering
    \includegraphics[width=\columnwidth]{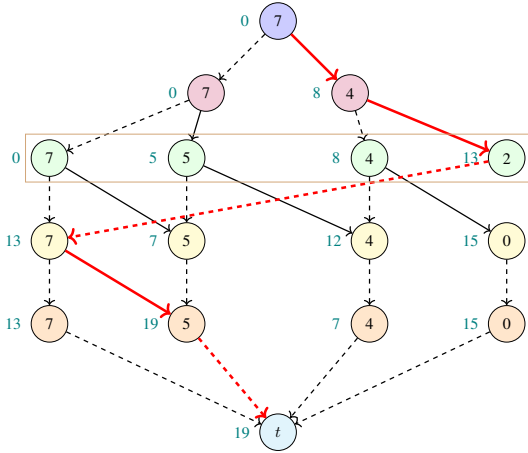}
    \caption{a Relaxed DD for \emph{Knapsack} problem \eqref{knapsack_eq}. The dashed arcs and the solid arcs correspond to the decisions of 0 and 1, respectively, for the decision variable. The \emph{longest} path is 19 (highlighted in red), obtained by traversing the path highlighted in red from the root $r$ to the terminal $t$.}
    \label{fig:relaxed_dd}
\end{figure}

Over the past decade, DDs have been adopted in discrete optimization, where they can encode the feasible region of integer programs. Their key advantage lies in their structured decomposition, where each layer of the tree corresponds to a decision variable, allowing partial solutions to be stored and reused. However, constructing an \emph{exact} DD will have exponential growth in both time and memory. Therefore, \emph{relaxed} and \emph{restricted} DDs are often used to maintain tractable widths while providing dual and primal bounds within branch-and-bound frameworks.  A \emph{restricted} DD (in Figure \ref{fig:restricted_dd}) is constructed by selecting only a subset of nodes of size \emph{max-width} and discarding the remaining nodes. The solutions in a restricted DD encode a subset of the original solution set of the \emph{exact} DD. In contrast, a \emph{relaxed} DD (in Figure \ref{fig:relaxed_dd}) is constructed by selecting a subset of nodes of size \emph{max-width} and \textit{merging} them into a single node with a relaxed state value that inherits all the arcs connected to these merged nodes. This node-merging process ensures that feasible solutions of the original set are encoded by a subset of the paths in the resulting DD and provides a superset of the original solution set. 

The {restricted} DD provides a \emph{primal} bound (lower-bound in the maximization problem) and the {relaxed} DD provides a \emph{dual} bound (upper-bound in the maximization problem) to the optimal solution. Iterative refinement of restricted and relaxed DDs through a branch-and-bound framework drives the primal–dual gap to zero.
By leveraging this structure, DD-based methods enable more efficient computation and can accelerate the branch-and-bound process. 
Recently, it has been shown that DD-based techniques can be extended to mixed-integer settings, where optimization problems involve both discrete and continuous variables, thereby significantly broadening the class of real-world problems that can benefit from this technology \cite{davarnia:2021, davarnia:va:2020, davarnia2024graphicalframeworkglobaloptimization}. Empirical studies further demonstrate that these methods can outperform state-of-the-art optimization solvers across a wide range of applications, including transportation, energy systems, and sparse regression in statistical learning \cite{salemi:davarnia:2022, salemi:da:2023, davarnia:ki:2025, davarnia:ri:ic:ta:2019, khademnia:da:2024}.

However, even these methods struggle with the scale of real-world problem instances. The encouraging factor is that DDs naturally lend themselves to parallelization due to their special graph-based structure \cite{bergman:ci:sa:sa:sa:va:2014}. This property can be exploited to enhance solution speed and make large-scale problems tractable.
Parallelization is particularly effective during the branch-and-bound process, where DD nodes are explored recursively to compute bounds that are further strengthened by incorporating cutting planes derived from suitable decompositions in mixed-integer settings \cite{salemi:davarnia:2022}. Since the search requires handling a vast number of nodes, each potentially producing further nodes, the branch-and-bound tree grows rapidly until convergence between primal and dual bounds is reached (yielding the optimal solution).
To address scalability challenges and reduce solution times, our research focuses on a key component for parallelizing the DD-based solver in a multicore environment, with the goal of efficiently distributing node exploration across multiple processors.

The DD-based solver employs a master-worker architecture in which each worker retrieves a node from its private queue and explores it \cite{hooker_book}. During this, the worker may generate additional nodes and push them back into the queue. When a worker's queue becomes empty, the master thread intervenes by stealing nodes from busy workers and reallocating them to the idle ones. In addition, the master maintains bookkeeping information about worker states and dynamically adjusts the allocation strategies to maintain balanced utilization.

\subsection{Contributions}
Our key contributions are as follows.
\begin{itemize}
    \item We introduce a new lock-free work-stealing queue tailored for master-worker architectures. This design supports native bulk operations, allows unbounded queue growth without resizing, and adopts a simplified concurrency model with a single stealer.
    \item We provide a conceptual proof sketch of correctness by identifying linearization points for each operation and showing that the algorithm preserves safety while guaranteeing lock-free progress.
    \item We present extensive benchmarks and a pseudo workload study demonstrating that the proposed queue achieves constant-latency performance for both bulk pushes and bulk steals. We further show an optimization that can significantly reduce steal latency in common-case scenarios.
\end{itemize}

The remainder of the paper is organized as follows. Section ~\ref{sec:motivation} outlines the requirements that shaped our design. Section \ref{sec:algorithm} presents the proposed algorithm, and Section ~\ref{analysis} provides the correctness arguments. Section ~\ref{sec:results} reports our experimental evaluation. Section \ref{sec:related_work} reviews prior work on work-stealing queues. Finally, Section \ref{sec:conclusion} concludes the paper with limitations and directions for future research.

\section{Motivation}\label{sec:motivation}
\subsection{Requirements}


In designing a work-stealing algorithm tailored to the parallel version of the DD-BD solver \cite{salemi:da:2023} within a master-worker framework, we identified several requirements that are not fully addressed by existing algorithms. These requirements arise from the specific workload characteristics of our solver as well as scheduling policies enforced by the master process. The key requirements are summarized below:

First, the queue should support bulk operations. In our setting, new node generation typically occurs in bulk, often producing more than a hundred nodes at once. Similarly, when the master attempts to acquire nodes through stealing, it is preferable to obtain a proportion of available nodes in the queue in a single operation rather than one at a time. Supporting bulk push and bulk steal operations reduces the waiting time of the workers, as multiple tasks can be transferred in a single function call, and it also reduces overall steal and push attempts in the system. While it is possible to simulate bulk operations by wrapping repeated single-node operations, such an approach introduces unnecessary overhead. Therefore, native support for bulk operations is a fundamental requirement.

Second, the queue must be unbounded in order to accommodate workloads that generate a very large number of nodes. At the start, the queue can grow rapidly before gradually shrinking as nodes are explored. Fixed-size array implementations are unsuitable in this context, as the information about the array capacity cannot be predetermined at compile time. Similarly, dynamically resizing arrays are also undesirable because they incur expensive element-wise copying during expansion or contraction. Consequently, the queue must inherently support unbounded growth, subject only to memory limitations. 

Third, at any given time, a queue is updated by at most two concurrent threads: the owner and a single stealer. The number of concurrent stealers is limited to one, because the master acts as a global load balancer, and is solely responsible for stealing and assigning nodes to idle workers as per the scheduling policy. This constraint simplifies the synchronization requirements and stopping criteria of the solver, and allows for a more efficient design \cite{satx10}.

Fourth, consistent with the modern work-stealing designs, we aim to avoid the use of locks. Given that at most two concurrent threads update a queue, we do not require the strict guarantees of serializable operations provided by the lock-based synchronization. Moreover, avoiding mutexes eliminates the associated synchronization overhead, which is particularly costly in fine-grained parallel settings. Instead, progress can be ensured through non-blocking synchronization primitives such as atomic operations. 

Together, these requirements form the basis for the design of our new algorithm, which we present in the following section.

\subsection{Master-Worker model}
The choice of master-worker architecture with a single designated stealer arises naturally from the characteristics of the DD-BD solver that we currently work on to parallelize. During node exploration, along with child nodes, a worker generates cuts, which are constraints that tighten the feasible region and enable aggressive pruning of infeasible and suboptimal solutions. These cuts are most effective when applied to nodes originating from the same or closely related regions of the search space, as such nodes share similar structural properties \cite{laurent:joxan, moisan:2013, local:search}.

To exploit this locality, the solver maintains a centralized view of the search process. The master maintains global bookkeeping, tracking which subspaces are currently explored by each worker, monitoring load statistics, and coordinating adaptive load balancing decisions. This responsibility belongs to the scheduling policy rather than to the queue abstraction itself, since it depends on solver-specific runtime information such as subspace affinity, pruning effectiveness, and workload evolution. The work-stealing queue provides a mechanism for transferring tasks efficiently, while the master determines when, from whom, and to whom work should be reassigned.

When load imbalance arises, the master selectively steals batches of nodes from busy workers and redistributes them to idle ones. Importantly, the master can identify subsets of nodes that are topologically closest to a worker’s current subspace and assign them accordingly, preserving locality and maximizing the effectiveness of previously generated cuts. Allowing multiple concurrent stealers would undermine this strategy by introducing contention at worker queues and randomizing task distribution, thereby reducing the locality benefits that are crucial for solver performance.

A centralized master also facilitates future extensions of the solver. In particular, we plan to support distributed execution using MPI, where a single coordinator with complete bookkeeping simplifies global scheduling decisions and data movement across machines. These extensions are part of ongoing work, but they further motivate the design choice of a single master responsible for stealing and redistribution.

The master’s primary responsibility is to coordinate load balancing and maintain a global view of the search. Although the single-stealer design may appear restrictive, it reflects a deliberate tradeoff between scheduling flexibility and global coordination. By centralizing stealing, the master can make informed redistribution decisions based on the overall search state. Importantly, the master does not permanently occupy a dedicated core; when all the workers are busy, it participates in  node exploration and initiates stealing only when workers become idle.

The master also deliberately waits until a worker’s local queue is nearly drained before redistributing work. Individual nodes can cascadingly generate hundreds of new nodes, particularly when they correspond to promising or optimal solutions. Premature redistribution would risk assigning work that would soon need to be stolen back after such expansion, increasing unnecessary steal operations and disrupting locality. This design ensures that redistribution occurs only when it is likely to be stable and beneficial.

This behavior of master ensures that the master contributes computational work whenever possible, while intervening only when global load imbalance arises.

\section{Algorithm}\label{sec:algorithm}

The dequeue is implemented with a singly linked list and maintains two atomic variables \lstinline{size} which corresponds to the number of elements currently in the queue and \lstinline{head} pointer that points to the current head of the queue. Both these values are updated in each API operation to ensure correctness under concurrent access. Each node is padded with additional memory to prevent false sharing. 
Though this data structure is labelled as deque, all the traversals begin with the head node, and ends on seeing a null pointer that marks the end of the list. The class for deque is shown in Listing ~\ref{class}. 

\begin{lstlisting}[language=c++, float, label=class, caption=Simplifed pseudocode of our Queue class and its API operations]
class lf_node {
public:
  uint8_t padding[CACHE_LINE_SIZE];
  lf_node *next;
};

typedef tuple<lf_node*, lf_node*, size_t> llist;

class lf_queue {
private:
  std::atomic<size_t> size = 0;
  std::atomic<lf_node *> head = nullptr;
public:
  llist steal(double proportion);
  lf_node *pop();
  void push(const llist &nodes);
};
\end{lstlisting}

\subsection{API operations}\label{sec:api}

\emph{\textbf{push}} : (Listing \ref{push}): Given a batch of nodes, as a linked list, this function atomically pushes the new nodes to the head of the queue. It links the 'end' of the incoming list to the current head of the queue, then atomically updates the queue's \lstinline{head} pointer to point to the start of the new list. Finally, it atomically increments the size of the queue by the number of nodes added. In-depth analysis of memory order semantics is provided in Section ~\ref{analysis}.

\begin{lstlisting}[language=c++, float, caption=Push operation invoked by the owner to insert given batch of nodes at the head of the queue, label=push]
void push(const llist &nodes) {
  lf_node *start = nodes.start, *end = nodes.end;
  size_t n = nodes.n;
  end->next = head.load(RELAXED);
  head.store(start, RELEASE);
  size.fetch_add(n, ACQUIRE_RELEASE);
}
\end{lstlisting}

\emph{\textbf{pop}} : (Listing \ref{pop}), called by the owner to extract and remove the top element (head) of the queue and effectively updates the head and size atomically. This return a null pointer when the queue is empty.

\begin{lstlisting}[language=c++, float, label=pop, caption=Pop operation invoked by owner to remove and return the head of the queue.]
lf_node *pop() {
  lf_node *rv = head.load(RELAXED);
  if(!rv) return NULL;
  head.store(rv->next, RELAXED);
  size.fetch_sub(1, ACQUIRE_RELEASE);
  rv->next = NULL;
  return rv;
}
\end{lstlisting}

\begin{figure}
    \centering
    \includegraphics{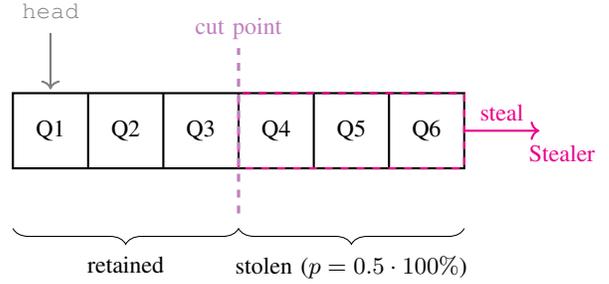}
    \caption[Bulk Stealing]{Bulk stealing: the stealer removes a proportion $p$ (dashed region) from the tail (right) in a single operation. The queue is severed at the \emph{cut point} (dashed line), after which the suffix becomes the stolen sublist. Example shown with $p=50\%$.}
  \label{fig:bulk_steal_cut_long}
\end{figure}

\emph{\textbf{steal}}: (Listing \ref{steal}): The function is invoked by the stealer to detach and acquire a contiguous sublist of nodes from the tail of the queue. Given a \lstinline{proportion} parameter, the function determines a cut point (as shown in Figure \ref{fig:bulk_steal_cut_long}) such that approximately the specified fraction of nodes remains in the queue after the operation, and the remaining suffix is removed and returned as a linked list segment. The operation begins by reading the current size of the queue and computing the number of nodes to skip from the head to reach the cut point. It traverses the list to reach the cut point, performs a consistency-check to ensure that the queue has not drained significantly, records the \lstinline{next} pointer and severs the list by setting the \lstinline{next} pointer of the cut node to \lstinline{null}. The detached segment is traversed to identify its tail and to count the number of stolen elements, after which the \lstinline{size} counter is atomically decremented by this amount. If the queue is too small, or if consistency-check reveals rapid draining, the steal attempt is aborted and no nodes are returned.

\begin{lstlisting}[language=c++, float, label = steal, caption=Pop called by stealer to steal a given proportion of nodes from the tail of the queue. If successful this returns a batch of nodes as a linked list.]
llist steal(double proportion) {
  // % of nodes to be left after this operation.
  proportion = 1 - proportion; 
  size_t sz = size.load(ACQUIRE);
  if (sz < _queue_limit_) return {NULL, NULL, 0};

  size_t n_skip = sz * proportion;
  size_t rem = sz - n_skip; // # nodes to be popped from the end.
  size_t k = n_skip;

  lf_node *start = head.load(ACQUIRE);
  while (n_skip && start) {
    start = start->next;
    n_skip--; 
  }

  if (n_skip || !start) 
      return {NULL, NULL, 0}; // not enough nodes.

  size_t ssz = size.load(ACQUIRE);
  if (ssz <= (sz - (k>>1))) 
      // nodes are popped very quick, abort.
      return {NULL, NULL, 0}; 

  lf_node *begin = start->next;
  start->next = nullptr;
  size.fetch_add(0, RELEASE);

  // reach the tail of the list and update size.
  lf_node *end = begin;
  size_t count = 0;
  while (end) {
    count++;
    if (!end->next) break;
    end = end->next;
  }
  size.fetch_sub(count);
  return {begin, end,count};
}
\end{lstlisting}

\subsection{Analysis }\label{analysis}

The \lstinline{push} function is invoked exclusively by the owner thread. It begins with a \emph{relaxed} atomic load of the \lstinline{head} pointer. Because the \lstinline{head} is only ever modified by the owner, a relaxed load suffices: it is guaranteed to observe the most recent value written by the same thread in its prior \lstinline{push} or \lstinline{pop} operation \cite{cpp_cia} \footnote{No other thread updates \lstinline{head}, so there is no risk of observing a stale or inconsistent value. The \emph{relaxed} load simply reuses the last write. }. The subsequent store to \lstinline{head} is performed with \emph{release} semantics. This ensures that all preceding memory instructions in the program order become visible before the update to \lstinline{head}, preventing earlier instructions from being reordered past the store.

The update to \lstinline{size} counter is carried out atomically with \emph{acquire-release} semantics \footnote{An \emph{acquire-release} read-modify-write instruction prevents both forward and backward reordering: prior instructions cannot move past the update, and subsequent instructions cannot move before it. }. This choice enforces a bidirectional ordering: prior instructions cannot be moved after the update, and subsequent instructions cannot be moved before it. As a consequence, any other thread that performs a load of \lstinline{size} with \emph{acquire} semantics and observes this update is guaranteed to see the corresponding new value of \lstinline{head}. In effect, the \emph{acquire-release} operation acts as a memory barrier, ensuring consistency between the modifications to \lstinline{head} and \lstinline{size}.

The correctness of the \lstinline{steal} operation follows from the structural invariants of the queue and the restricted concurrency model. At most two threads interact with the queue concurrently: the owner and a single stealer. The queue is a singly linked list rooted at \lstinline{head}, with the end marked by a \lstinline{null}. The owner performs pushes by prepending at the \lstinline{head} and pops by advancing the \lstinline{head}, but never rewires internal \lstinline{next} pointers. The stealer traverses the list from a snapshot of the \lstinline{head} (read with \emph{acquire} semantics) to locate the cut position, then severs the list at the cut point (Line 26). This single write defines the linearization point: after it, the prefix remains reachable from the head, and the suffix is returned to the stealer. Since the owner never reconnects internal links, and only prepends or shortens the prefix, no suffix node can be duplicated or lost. 

The code includes multiple consistency checks to abort the operation if the queue changes too rapidly during traversal. Lines 20-21 (Listing ~\ref{steal}) detects aggressive draining by reloading the current size and verifying that at least half of the initially observed nodes remain. These checks prevent the stealer from cutting into a queue that has been nearly emptied by the owner. The store (Line 27) performed with \emph{release} semantics ensures that the update, setting the \lstinline{next} to \lstinline{null} (Line 26) is properly published, so that subsequent push and pop operations by the owner observe the cut consistently. Together with the atomic decrement of the \lstinline{size}, this guarantees that once the operation returns, no stolen node remains reachable from the head. Thus, every node is either retained in the prefix or returned in the stolen suffix, with the size updated accordingly. The operation is lock-free, since the stealer either succeeds with a single cut and updates the size or aborts immediately when conditions are not satisfied, ensuring that at least one thread always makes progress. 

The operations of our algorithm are linearizable. Each operation admits a well-defined linearization point between its invocation and response. a \lstinline{push} linearizes at the write that installs the new head pointer, a \lstinline{pop} linearizes at the update advancing the head to its successor, and a \lstinline{steal} linearizes at the write that serves the list by setting \lstinline{start->next=null}. These points ensure that real-time order is preserved: pushes and pops that complete before a steal begins are visible to the stealer through acquire loads, while steals that complete before subsequent operations are observed consistently due to release semantics. From an external perspective, the sequence of operations is indistinguishable from a valid sequential execution of a deque where the owner inserts and removes tasks at one end and the stealer removes tasks from the other. Thus, the implementation satisfies linearizability without requiring locks.

\section{Evaluation} \label{sec:results}
 
We evaluated the performance of our algorithm (LF\_Queue) both by benchmarking each API operation against two other work-stealing algorithms (Unbounded Queue, and Bounded Queue) from C++ Taskflow's library. The benchmark measures the latency of each API operation: push, pop, and steal in isolation. The queue state is reset after every iteration of the benchmark to ensure consistency across iterations. Hardware and Software configuration is shown in Table~\ref {tab:hardware_info}.

\begin{table}[htbp]
    \centering
    \renewcommand{\arraystretch}{1.15} 
    \resizebox{0.4\textwidth}{!}{%
    \begin{tabular}{|l|l|}
        \hline
        \textbf{Component}        & \textbf{Specification} \\ \hline
        Processor model           & Intel Xeon Platinum 8358 \\ \hline
        Base clock rate           & 2.60 GHz \\ \hline
        Sockets                   & 2 \\ \hline
        Cores per socket          & 32 \\ \hline
        Total cores               & 64 \\ \hline
        Hardware threads          & 128 \\ \hline
        L1 cache per core         & 48 KB \\ \hline
        Compiler                  & GCC 15 \\ \hline
        Language standard         & C++20 \\ \hline
        Compiler flags            & \texttt{-O3 -march=native} \\ \hline
    \end{tabular}%
    }
    \caption{Hardware and Software configuration used in our experiments.}
    \label{tab:hardware_info}
\end{table}

\begin{figure}[ht]
    \centering
    \includegraphics[width=\columnwidth]{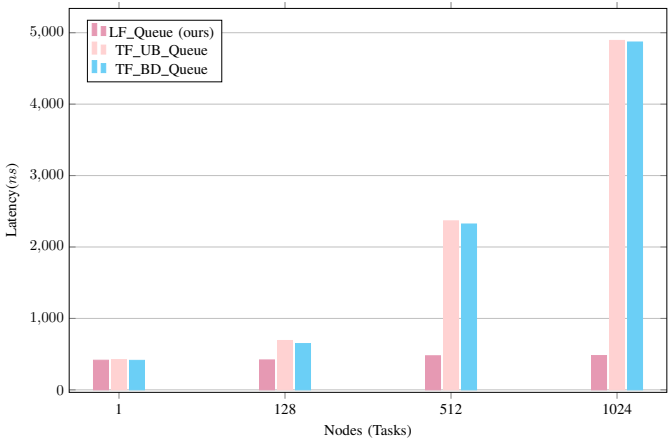}
    \caption{Average latency of \texttt{push} operation as a function of batch size. The proposed queue (LF\_Queue) maintains nearly constant latency regardless of the number of tasks pushed in the batch. In contrast, the Taskflow bounded (TF\_BD\_Queue) and unbounded queues (TF\_UB\_Queue) exhibit increasing latency as batch size grows.}
    \label{fig:push_ops}
\end{figure}

Figure ~\ref{fig:push_ops} shows the latency of the push operation as a function of batch size (1,128,512,1024). For small pushes (1-128 nodes), all three implementations perform similarly. However, as the batch size increases, the performance diverges. In both TF\_BD\_Queue and TF\_UB\_Queue, latency rises sharply with the number of nodes pushed, reaching nearly 5000 ns for 1024 nodes due to repeated per-node operations and resizing overheads. By contrast, LF\_Queue exhibits almost constant latency across all batch sizes, remaining below 500 ns even for 1024 nodes. This shows that LF\_Queue effectively amortizes the cost of bulk pushes and provides predictable performance independent of batch size, which is particularly advantageous in our workload where hundreds of nodes are generated at once.

\begin{figure}[ht]
    \centering
    \includegraphics[width=\columnwidth]{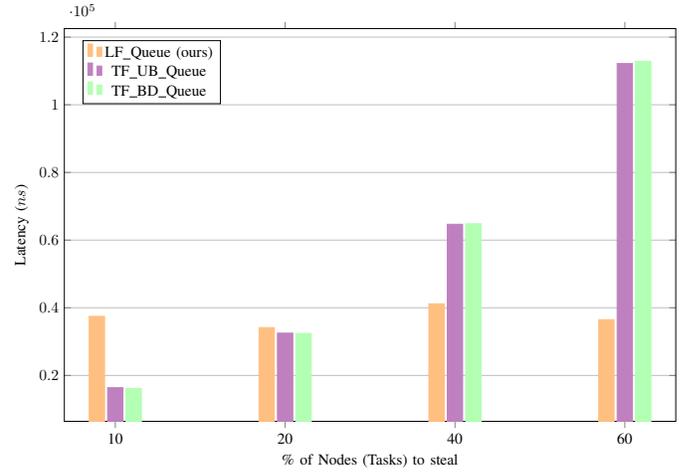}
    \caption{Average latency of \texttt{steal} operation as a function of the proportion of tasks stolen from a queue of initial size 10,000. 10\% indicates stealing 10\% of nodes from the tail of the queue with a single invocation of \lstinline{steal}. The proposed algorithm (LF\_Queue) shows higher latency than Taskflow (TF\_UB\_Queue and TF\_BD\_Queue) for smaller proportions (< 30\%) and performed better than other two implementations for larger proportions. }
    \label{fig:steal_ops}
\end{figure}

The results for steal operations are shown in Figure ~\ref{fig:steal_ops}. Here, the x-axis represents the proportion of nodes stolen from the queue. For example, a value of 20 indicates stealing 20\% of the tail of the  queue. With small steals(10\%), Taskflow's bounded and unbounded queues outperform LF\_Queue ($\approx 16 \mu s$ vs $\approx 39 \mu s$). However, as the proportion of stolen nodes increases, Taskflow's latency grows almost linearly, reaching over $112 \mu$s  when stealing 60\% of the queue. In contrast, LF\_Queue maintains stable performance, with latency consistently around $40 \mu$s regardless of the proportion stolen. This demonstrates the efficiency of LF\_Queue bulk stealing mechanism, which amortizes traversal costs over multiple nodes and avoids repeated synchronization.

A small optimization can be made to the \lstinline{steal} operation. In the original version, the stealer traverses the entire queue from the head to the tail, even after severing the list at the cut point. This second traversal is necessary to compute the exact number of stolen nodes, since the owner may push or pop concurrently and alter the queue size. In contrast, the optimized variant skips this second traversal when it can be guaranteed that the owner did not update the queue during the steal. In this case, the number of stolen nodes can be determined from the initial size and cut position, and the operation can return immediately after the list is severed.

\begin{figure}[ht]
    \centering
    \includegraphics[width=\columnwidth]{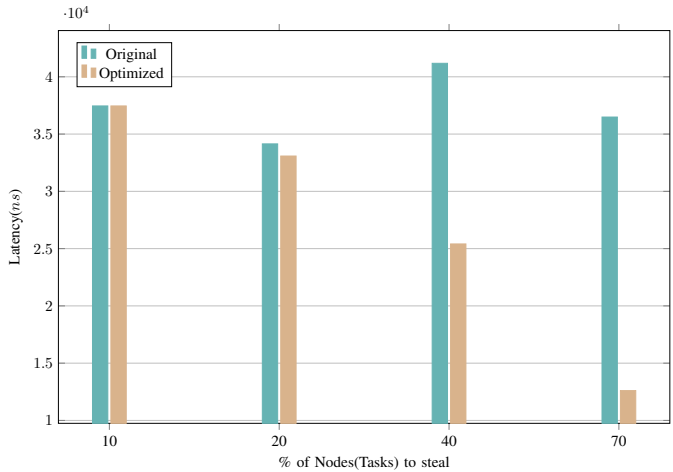}
    \caption{Comparison of average latency between original (\emph{non-optimized}) and optimized versions of \texttt{steal} operation in our algorithm. The optimized version avoids traversing to the end when the owner does not update the queue during the steal. This optimization significantly reduces latency (by 3x) as the proportion of stolen tasks increases.}
    \label{fig:opt_steal_ops}
\end{figure}

Figure ~ \ref{fig:opt_steal_ops} shows the latency of steal operations under both implementations. The non-optimized version requires approximately 37-38 $\mu$s regardless of the proportion of the stolen, reflecting the cost of traversing the full list. By contrast, the optimized version exhibits significantly lower latencies as the proportion increases: from 37.7 $\mu$s at 10\%  down to 12.5 $\mu$s at 60\%. This improvement arises because the optimized version avoids the linear tail traversal, amortizing the cost of stealing across larger proportions of the queue.

Although this case arises frequently in our target workloads, where workers typically remain busy processing their nodes concurrently with the master, we didn't integrate it into the baseline algorithm. We aimed to measure the latency of the \lstinline{steal} in the worst case, where traversal is unavoidable. Nevertheless, these results indicate that in practice, this optimization can substantially reduce steal latency in common scenarios.

For pop operations, all three queues perform similarly, with an average latency of $\approx 213-216 n$s, indicating no meaningful performance difference(figure omitted).

In addition to the microbenchmarks, we evaluated our queues under a more realistic workload. Ideally, this would involve our parallel solver; however, the solver is still under development, and its workload is difficult to simulate due to randomization and the inner subproblem solver. As an approximation, we implemented a pseudo workload based on the parallel exploration of a large directed acyclic graph (DAG). In this setup, each worker maintained a private work-stealing queue and began with an initial node. Outgoing nodes discovered during exploration were pushed to the worker's queue, provided that they had not already been claimed by another worker. When a worker's queue became empty, it attempted to steal half of the nodes from another worker's queue, selecting victims in order of worker IDs. To simulate the single concurrent stealer, each queue maintains an atomic flag that a stealer can toggle before and after stealing from this queue.

\begin{figure}[ht]
    \centering
    \includegraphics[width=1.0\columnwidth]{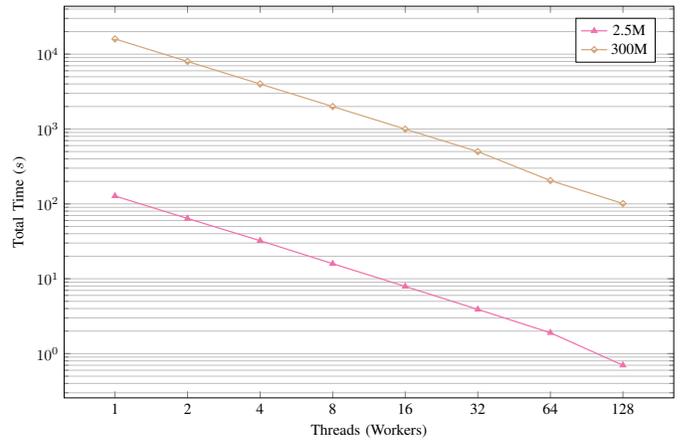}
    \caption{Execution time of DAG exploration using our queue with two input graph sizes (2.5M, 300M nodes) as the number of threads increases from 1 to 128. Both the x-axis and the y-axis are shown in logarithmic scale to emphasize the near log-linear scalability: execution time roughly halves when the number of threads doubles. }
    \label{fig:scalability_plot}
\end{figure}

We constructed two input graphs of sizes 2.5M nodes, and 300M nodes. Each experiment was executed with a varying number of threads, from 1 up to 128, and the total execution time was measured. The results, summarized in Figure ~\ref{fig:scalability_plot}, show consistent scalability of our algorithm across both the input graphs: the execution time decreases approximately by half when the number of threads is doubled, indicating near log-linear speedup. This trend hold uniformly across all input sizes, suggesting that the queue maintains efficiency even under substantially larger workloads. To visualize this behaviour, we present the results on a logarithmic scale. The two series corresponding to the different graph sizes exhibit identical slopes, highlighting that the scalability of the algorithm is not sensitive to the input graph size.

\begin{figure}
    \centering
    \includegraphics[width=\columnwidth]{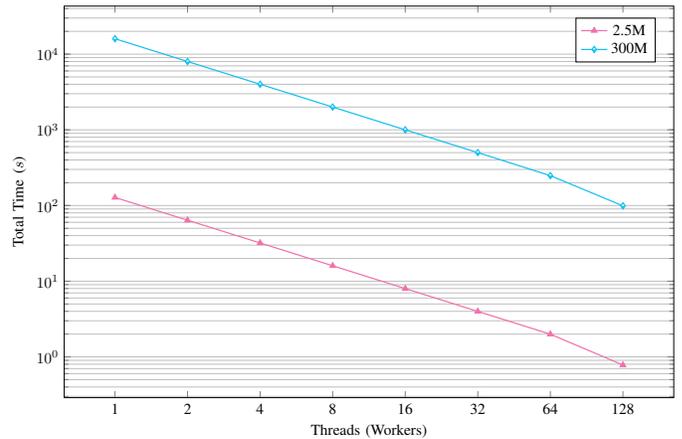}
    \caption{Execution time of DAG exploration using TF\_UB\_Queue with two input graph sizes (2.5M, 300M nodes) as the number of threads increases from 1 to 128. Both the x-axis and the y-axis are shown in logarithmic scale. Similar slope is observed for TF\_BD\_Queue and thus ignored.}
    \label{fig:scalability_tfub}
\end{figure}

Figure \ref{fig:scalability_tfub} shows the performance of the TF\_UB\_Queue in the above experiments. TF\_BD\_Queue also showed very similar performance and thus not included in the plots to preserve visibility. 
These results suggest that in this pseudo workload, the cost of individual queue operations does not dominate overall performance. Instead, scalability is primarily governed by the inherent parallelism in the DAG structure and the distribution of exploration tasks.

It is important to note, however, that in this DAG exploration benchmark, the cost of processing each node is uniform. In contrast, node processing times in our solver vary very high, and depend on various factors such as solution feasibility, undergoing pruning and refining, and solving a stochastic subproblem (dual linear program) by an inner solver. Such irregular workloads are more likely to amplify the impact of queue overheads, particularly under heavy contention. We therefore expect that once integrated into the solver, the benefits of our algorithm's bulk operations and simplified concurrency model will become more evident in reducing scheduling overhead.

\section{Related Work}\label{sec:related_work}

Work stealing emerged as a principled strategy for scheduling dynamic multithreaded computations with tight guarantees on time and space. The classical analysis by Blumofe and Leiserson \cite{blumofe} shows that randomized work stealing queues executes a computation with work $T_1$ and span $T_{\infty}$ in expected time $T_1/P + O(T_\infty)$ on $P$ processors, while using space within a constant factor of sequential space for fully-strict computations. This result established work stealing as both efficient and robust for nested parallelism, forming the theoretical basis for systems such as Cilk.

To make work stealing practical, efficient concurrent deques are required so that owners push/pop at one end while thieves steal from the other with minimal contention. Early work by Arora, Blumofe, and Plaxton \cite{arora_blumofe_plaxton} presented a nonblocking scheduler and deque interface in a multithreaded setting, establishing the importance of lock-free data structures for scalable scheduling. Hendler, Lev, and Shavit \cite{hendler_2004} extended this line by proposing dynamic work-stealing queues based on linked segments, enabling unbounded growth but introducing additional complexity. Chase and Lev \cite{chase_lev} later proposed the widely accepted dynamic circular work-stealing queue, a simple lock-free ring buffer that grows on demand and became the standard blueprint for runtime systems.

Work stealing underpins several mainstream runtimes. Lea's Java Fork/Join framework \cite{Java_forkjoin}, which later became part of the JDK, brought work-stealing into the Java ecosystem. Intel's Thread Building Blocks (TBB, now oneTBB) \cite{intel2019_tbb} adapted work stealing as its core task scheduling strategy in C++. Other systems, such as Go's runtime \cite{go_runtime} scheduler, also rely heavy on work stealing. These frameworks validated the generality and performance of work stealing in industrial settings, but they largely follow the Chase-Lev design and assume a general-purpose setting with multiple stealers and irregular workloads.

While Chase-Lev deques are correct under sequential consistency, they can fail on weak memory architectures such as ARM and POWER. L\^e, Pop, Cohen, and Zappa Nardelli \cite{cpp_tf_paper} addressed this gap by formally proving the correctness of Chase-Lev style deques under weak memory models. They identified the minimal set of acquire and release fences required to preserve linearizability and lock-freedom across architectures, while avoiding unnecessary synchronization overhead. Practical implementations derived from these designs include the C++ Taskflow library \cite{Huang2020Taskflow}, which provides bounded and unbounded lock-free queues. The bounded variant employs a ring buffer for predictable workloads, while the unbounded one supports dynamic growth for irregular ones. 

Beyond Chase-Lev, researchers have proposed alternative deque designs and scheduling strategies. For example, Lace \cite{lace_2014} introduced a split deque with a dynamic boundary between private and shared regions, reducing contention and minimizing fence usage. Other research explored adapting work stealing to DAG-based task graphs with dependencies and heterogeneous workloads. Taskflow \cite{Huang2020Taskflow}, for example, employs work stealing to schedule directed acyclic graphs of tasks while considering dependencies, reflecting a broader shift from flat task pools to graph based scheduling.

Locality and NUMA-awareness form another important direction. While work stealing is highly effective for load balance, it is agnostic to the memory hierarchy, and indiscriminate stealing can result in costly remote memory accesses on multisocket systems. Early analyses by Acar, Blelloch, and Blumofe \cite{acar2000} demonstrated that work stealing can provide good cache performance in expectation, but subsequent systems explicitly introduced locality-aware policies. Chen and Guo \cite{chen_2015} presented Locality-Aware Work Stealing, incorporated runtime profiling to adapt victim selection for multicore systems.  These contributions show that preserving locality while balancing load is both possible and beneficial, and that the placement and movement of tasks across cores can substantially influence performance.

To amortize scheduling overhead for bulk task operations, researchers have explored batch stealing techniques that allow a thief to grab multiple tasks in one attempt. A well-known strategy is steal-half policy, where a theif takes approximately half of the victim's queued tasks instead of just one. This approach, introduced by Hendler and Shavit \cite{steal_half_shavit}, reduces steal frequency and is used in modern runtimes like Go and Rust's Tokio to improve throughput. Batched stealing can signficantly lower synchronization per task, but it risks over-stealing: taking more work than needed, which can leave the victim thread underutilized and add overhead for moving excessive tasks. 
To address this, adaptive chunk sizing strategies have been proposed. Adnan and Sato \cite{ADNAN2012} developed a dynamic-length work-stealing algorithm that adjusts the number of tasks stolen based on the victim's load in a single-stealer scenario. Their strategy uses runtime information (e.g. the depth of the victim’s task stack) to determine an appropriate chunk size for each steal attempt. Compared to always stealing a fixed number of tasks, this adaptive approach avoids both under-stealing and over-stealing. 

A recent advancement along similar lines is Block-based Work Stealing (BWoS) \cite{bwos}, which generalizes the idea of segmenting a deque into multiple chunks or blocks. In BWoS, each per-core task queue is composed of several fixed-size blocks with independent metadata. The owner thread and thieves synchronize only at the level of whole blocks, rather than individual tasks. 
By stealing from the middle, BWoS avoids the thief always contending for the most recently produced tasks; instead, older blocks can be taken wholesale with minimal coordination.  Domain-specific extensions of work-stealing also arise, such as in  parallel garbage collection and for heterogeneous CPU-GPU systems. Adaptive schedulers such as SLAW \cite{slaw2010} switch between work-first and help-first strategies to optimize for both overhead and stack usage depending on program state. These efforts highlight the flexibility of work stealing as a general paradigm that can be tuned for diverse workloads and hardware architectures.

Our solver departs from general-purpose parallel runtimes by adopting a centralized master-worker architecture with structured task patterns. Most existing lock-free work-stealing algorithms, such as Chase-Lev, assume single-task operations, multiple concurrent stealers, which are mismatched with our solver's needs. Specifically, our workload involves bulk task generation and transfer, which existing algorithms simulate inefficiently through repeated single-node operations. We also enforce a single-stealer concurrency model, allowing us to simplify synchronization and reduce atomic overhead. Moreover, we require unbounded queue growth without resizing, which ring-buffer-based or dynamically resized dequeus fail to support efficiently.

Though designs like BWoS and Lace, are effective in highly concurrent general-purpose settings, they introduce complexity (e.g., block metadata, split-boundary managment) that is unnecessary in our two-threaded concurrency model. Additionally, their opportunistic and randomized stealing approaches do not align with our domain-aware, master-controlled task redistribution.  These constraints led us to develop a specialized queue with native bulk operations, simplified synchronization, and unbounded growth, that align with the needs of our decision diagram-based solver.

In our evaluation, we compared against both Taskflow implementations, as these implementations were readily available and integrated smoothly with our test harness. While other lock-free work-stealing algorithms such as Lace or BWoS offer open-source implementations, we encountered significant difficulty adapting them to our experimental setup, particularly due to differences in runtime assumptions and data structures. We showed that our algorithm, designed for a solver-specific workload, achieves competitive or superior performance by incorporating bulk operations, unbounded queue growth, and a simplified single-owner/single stealer concurrency model.

\section{Limitations and Conclusion} \label{sec:conclusion}

In this work, we presented a new lock-free work-stealing algorithm tailored to the requirements of our graph-based optimization solver. Unlike existing general-purpose algorithms, our queue supports bulk operations, and assumes at most one concurrent stealer. These design choices allow us to avoid costly synchronization, simplify correctness reasoning, and enable efficient implementations of push, pop, and steal operations.

Our evaluation shows that the proposed algorithm outperforms the state-of-the-art implementations in C++ Taskflow. Most notably, it achieves constant-latency push performance, in contrast to Taskflow’s queues, whose latency increases sharply with batch size. This yields predictable performance in scenarios where large batches of tasks are generated. Pop operations perform comparably across all implementations. For steals, our queue maintains stable latency even as the proportion of steal operations increases, whereas Taskflow queues degrade significantly. We also investigated an optimized steal variant that avoids full traversal when the owner is idle, reducing latency by up to 3x. Although excluded from the baseline to capture worst-case performance, this optimization is particularly relevant in practical workloads. Finally, in a pseudo workload based on DAG exploration, all three queues scaled linearly and achieved nearly identical completion times, reflecting the uniform cost of node processing. 

Our study also has limitations. First, while the DAG workload provides useful insight into scalability, it does not fully capture the irregular and heuristic-driven behavior of our target solver. In practice, node processing times during the solution process of the solver can vary widely, and we expect this variability to further amplify the benefits of bulk operations and simplified concurrency. However, fully integrating the algorithm into the solver remains future work. Second, our algorithm assumes a single concurrent stealer and does not directly support multiple concurrent stealers.

Despite these limitations, our results demonstrate that tailoring work-stealing algorithms to domain-specific characteristics can yield tangible benefits. Future work will focus on integrating the queue into the solver to validate its performance in real workloads, extending the design to support multiple concurrent stealers when required, and exploring further optimizations such as the early-return steal path.

Finally, we emphasize that our algorithm is not intended to outperform general-purpose work-stealing queues in all settings. In particular, we do not claim improved performance for small, fine-grained steal operations, nor for systems that permit many concurrent stealers per queue. Our design deliberately targets workloads with coarse-grained task generation, bulk stealing, and centralized scheduling, where simplifying synchronization and amortizing traversal costs are more beneficial than minimizing single-steal latency.

\bibliographystyle{ieeetr}
\bibliography{refs}

\end{document}